\documentstyle[multicol,aps,psfig]{revtex}

\begin{document}     



\title{New Optimization 
Methods for Converging Perturbative Series with a Field Cutoff}
\author{B. Kessler, L. Li and Y. Meurice
{\footnote {e-mail:yannick-meurice@uiowa.edu}}  \\
{\it Department of Physics and Astronomy, The University of Iowa, 
Iowa City, Iowa 52242, USA}}

\maketitle
\begin{abstract}
We take advantage of the fact that in $\lambda \phi ^4$ problems,  
a large field cutoff $\phi _{max}$
makes perturbative series {\it converge} toward values exponentially
close to the exact values, to make optimal choices of $\phi _{max}$.
For perturbative series terminated at even order, it is in principle
possible to adjust $\phi _{max}$ in order to obtain the exact
result. For perturbative series terminated at odd order, the error can
only be minimized. It is however possible to introduce a mass shift 
$m^2\rightarrow m^2(1+\eta)$ in order to obtain the exact
result. We discuss weak and strong coupling methods to determine $\phi
_{max}$ and $\eta$. The numerical calculations in this article have
been performed with a simple integral with one variable. We give
arguments indicating that the qualitative features observed should
extend to quantum mechanics and quantum field theory. We found that 
optimization at even order is more efficient that at odd order.
We compare our methods with the linear $\delta$-expansion (LDE)
(combined with the principle of minimal sensitivity) which
provides an upper envelope of for the accuracy curves of various
Pad\'e and Pad\'e-Borel approximants. Our optimization method performs
better than the LDE at strong and intermediate
coupling, but not at weak coupling where it appears less robust and
subject to further improvements. We also show that it is possible to 
fix the arbitrary parameter appearing in the LDE
using the strong coupling expansion, in order to get accuracies
comparable to ours. 
\end{abstract}

\begin{multicols}{2}\global\columnwidth20.5pc
 
\section{Introduction}

Perturbative methods in quantum field theory and their graphical
representation in terms of Feynman diagrams  can be 
credited for many important physics accomplishments of the 20th century 
\cite{feynman85,sirlin99}. 
Despite these successes, it is also well-known that 
that perturbative series are asymptotic \cite{dyson52,leguillou90}. 
In concrete terms, this means that 
for any fixed coupling, there exists an order $K$ in perturbation
beyond which higher order terms cease to provide a more accurate
answer. In practice, this order can often be identified by the fact  
that the $K+1$-th contribution becomes of the same order or larger than 
the previous ones. The ``rule of thumb'' consists then in dropping 
all the contribution of order $K+1$ and larger, allowing errors that
are usually slightly smaller than the $K$-th contribution.

For low energy processes involving only electromagnetic interactions, 
the rule of thumb would probably be satisfactory. On the other
hand, when electro-weak or strong interactions are turned on, it seems
clear that for some calculations the errors associated
with this procedure are getting close to the experimental error bars of
precision test of the standard model \cite{ym02}. In some cases, the 
situation can be improved by using Pad\'e approximants and/or Borel 
transforms\cite{baker96}. 
However such methods rarely provide rigorous error bars
and do not always work well at large coupling or when non-perturbative
effects are involved.

In the 21-st century, comparison between precise experiments and
precise calculations may become our only window on the physics beyond
the standard model. It is thus crucial to develop methods 
that go beyond the rule of thumb and provide controllable error bars
that can be reduced to a level that at least matches the experimental
error bars. In order to achieve this goal, we need to start with
examples for which  it is possible to obtain accurate numerical answers 
that can be compared with improved perturbative methods. 
This can be achieved with a reasonable amount of effort in the case of
scalar field theory (SFT), which we consider as our first target.

For SFT formulated with the path integral formalism, 
it has been established\cite{pernice98,ym02} that the large field
configurations are responsible for the asymptotic nature of the 
perturbative series. A simple solution to the problem 
consists in introducing a uniform large field cutoff, in other 
words, restricting the field integral at each site to $|\phi_x|<\phi_{max}$. 
This yields series converging toward values that are exponentially close to
the original ones\cite{ym02} provided that $\phi_{max}$ is large
enough. Numerical examples for three models \cite{ym02},
show that at fixed $\phi _{max}$, the
accuracy of the modified series peaks at some special value of the 
coupling. At fixed coupling, it is possible to
find an optimal value of $\phi_{max}$ for which the accuracy of the 
modified series is optimal. 
The determination of this optimal value is the main question discussed
in the present article.
When comparing the three subgraphs of
Figs. 2 and 3 of Ref. \cite{ym02} which illustrate these features, 
one is struck with the similarity 
in the patterns observed for the three models considered (a simple 
integral, the anharmonic oscillator and and SFT in 3-dimensions in the 
hierarchical approximation). It is thus reasonable to develop
optimization 
strategies with the simplest possible example, namely the one-variable
integral, for which the calculation of the coefficients of various
expansions does not pose serious technical difficulties. 
As we will see, there exist several ways to proceed and the
complicated dependence of the accuracy on the coupling constant 
certainly justifies this initial simplification.

In this article we address the question of the optimal choice of the 
field cutoff $\phi_{max}$ with the simple integral 
\begin{equation}
Z(\lambda)=\int_{-\infty}^{+\infty}d\phi {\rm e}^{-(m^2/2)\phi^2-\lambda 
\phi^4}\ .
\label{eq:int}
\end{equation}
This integral can be seen as a zero dimensional field theory. It 
has been often used to develop new perturbative methods 
\cite{leguillou90}, in particular the LDE
\cite{buckley93}. The coefficient of the quadratic term $m^2$ is
set to 1 in all the numerical calculations discussed hereafter, however
it will sometimes be used as an expansion parameter .

The effects of a field cut on this integral and 
the reason why it makes the perturbative series converge 
are reviewed in section \ref{sec:int}. Some useful features of the 
strong coupling expansion to be used later are discussed in section 
\ref{sec:sc}. 
Our treatment will be different for even and odd orders.
For series truncated at even orders, the overshooting
of the last positive contribution can be used to cancel the
undershooting effect of the field cut. In other words,
the errors due to
the truncation of the series and the field cutoff compensate exactly  
for a special value of the field cutoff $\phi_{max}^{opt}(\lambda)$.
This value is calculated approximately using weak and strong
coupling expansion in section \ref{sec:even}. 
For series truncated at odd orders (section \ref{sec:odd}), 
the two effects go in the same direction and the error can
only be minimized. However, an exact cancellation can be obtained 
by using a mass shift $m^2\rightarrow m^2(1+\eta)$. We then need to
find $\eta(\phi_{max},\lambda)$ such that the cancellation occurs.
In practice, it is desirable to have $\eta$ as small as possible and 
we will in addition impose that $\partial \eta/\partial \phi_{max}=0$.
This condition fixes the otherwise unspecified $\phi_{max}$.

The methods presented here have qualitative feature that can be
compared with the LDE \cite{buckley93}, where the 
arbitrary parameter can be seen as providing a smooth cut in field
space, or with
variational methods \cite{var} , where weak and strong coupling
expansions are combined.  
This is discussed 
in section \ref{sec:compa}. The main conclusion is that the 
method which consists in determining the value of $\phi_{max}$ which
is optimal for even series in the weak coupling, 
using the strong coupling expansion provide excellent results 
at moderate and strong coupling. We also show that it is possible to 
fix the arbitrary parameter appearing in the LDE
using the strong coupling expansion, in order to get accuracies
comparable to ours. 
In the conclusions, we discuss possible improvement at weak coupling 
and the extension of the model in more general situations.

\section{Effects of a field cutoff}
\label{sec:int}

In this section, we discuss the effects of a field cutoff for the 
integral defined by Eq. (\ref{eq:int}). We first discuss the 
problems associated with usual perturbation theory.
The basic question in ordinary perturbation theory is to decide for 
which values of the coupling, 
the truncated series at order $K$ is a good approximation, 
which in our example means
\begin{equation}
Z(\lambda)\simeq \sum_{k=0}^Ka_k\lambda^k\ ,
\end{equation}
with perturbative coefficients
\begin{eqnarray}
\nonumber
a_k&=&{(-1)^k\over{k!}}\int_{-\infty}^{+\infty}d\phi \  
{\rm e}^{-(m^2/2)\phi^2}
( \phi^4)^k\\
&=&{(-1)^k\over{k!}}\Gamma(2k+1/2)(2/m^2)^{2k+1/2}\ .
\label{eq:asubk}
\end{eqnarray} 
The ratios $a_{k+1}/a_k\simeq -16k$ grow linearly 
when $k\rightarrow \infty$ and in order to get a good 
accuracy at order $K$, we need to require 
$\lambda<<1/16 K$. 

An alternate way of seeing this is that the
integrand 
${\rm e}^{-(m^2/2)\phi^2}\phi^{4k}/k!$ is maximum at 
$\phi=2\sqrt{k}$. On the other hand, 
the truncation of 
${\rm e}^{-\lambda \phi^4}$ at
order $K$ is accurate provided that $\lambda \phi^4 <<K$.
The truncated expansion of the exponential is a good approximation
up to the region where the integrand is 
maximum, provided that 
$\lambda (2\sqrt{K})^4 <<K$,
which implies
$\lambda<<1/16 K$. 

\begin{figure}
\centerline{\psfig{figure=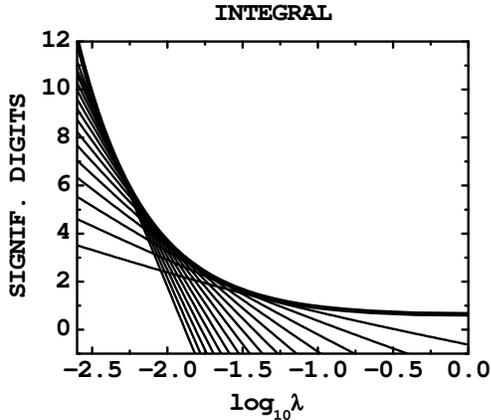,width=3.in}}
\caption{Number of significant digits versus $\lambda$ at order 1, 2,
  ... 15. In all graphs, all the logs are in base 10.
As the order increases, the curve rotates clockwise and moves
left. The thick line is the envelope Eq. (\ref{eq:env}). The large
blank area in the upper right corner is the region not accessible in
regular perturbation theory. }
\label{fig:ptenv}
\end{figure}

It is useful to represent the above discussion graphically. The number of 
of significant digits as a function of the coupling is given in
Fig. \ref{fig:ptenv}. 
It is important for the reader to get familiar with this kind of
graph, because we will use them in multiple occasions later in the paper.
The number of significant digits is $minus$ the
log in base 10 of the relative error. At sufficiently small coupling, 
the behavior becomes linear with a slope which is minus the order. 
Remembering the minus sign above, the intercept diminish with the
order. It is possible to construct an envelope for 
the curves at various order, in other words, a curve that 
lays above all the curves and is tangent at the point of contact. 
In Fig. \ref{fig:ptenv}, we have used a semi-empirical formula to draw
an approximate envelope: we have used the order $k$ as the
(continuous) parameter of a parametric curve 
\begin{eqnarray}
\label{eq:env}
x&=&-{\rm Log}_{10}(16 k)\\ \nonumber
y&=& -{\rm
  Log}_{10}\left[\left(16(k+1)\right)^{-k-1}|a_{k+1}|\pi^{-1/2}\right]
\end{eqnarray} 
A careful examination of the figure at low coupling (e. g., near 
$10^{-2})$ shows that as the order increases, the accuracy increases up
to an order where it starts decreasing. 
The envelope is the boundary to a range of accuracy that is
inaccessible using ordinary perturbation theory.

A simple way to convert the asymptotic series into a converging 
one \cite{pernice98,ym02}
consists in restricting 
the range of integration  
to $|\phi|<\phi_{max}$. On the restricted domain, 
${\rm e}^{-\lambda \phi^4}$ converges
uniformly and one can then interchange legally the sum and the integral.
We are then considering a modified problem
namely the perturbative evaluation of
\begin{equation}
Z(\lambda , \phi_{max}) =\int_{-\phi_{max}}^{+\phi_{max}}d\phi \  {\rm e}^{-(m^2/2)\phi^2-\lambda 
\phi^4}\ .
\label{eq:cutint}
\end{equation}
As the order 
increases, the peak of the
integrand of $a_k$ (see Eq. (\ref{eq:asubk}))
moves across $\phi_{max}$ and the large order coefficients
are suppressed by an inverse factorial: $|a_k|<\sqrt{2\pi}\phi_{max}^{4k}/k!$.
At the same time, we have an 
exponential control of the error: 
\begin{equation}
|Z(\lambda)-Z(\lambda , \phi_{max})|<2{\rm e}^{-\lambda \phi_{max}^4}
\int_{\phi_{max}}^{\infty}d\phi \ {\rm e}^{-(m^2/2)\phi^2}\\ 
\end{equation}

Everything works in a very similar way for other numerically solvable
$\lambda \phi ^4$ problems in $D=1$ (anharmonic oscillator) and $D=3$ 
(scalar field theory in the hierarchical approximation). The only
difference being that in these two cases, a more demanding
computational effort is required. This should be kept in mind 
while discussing the general strategy to be followed.
If we could calculate as many perturbative coefficients as needed, an 
obvious strategy would be to pick a field cut $\phi_{max}$ 
large enough to satisfy 
some accuracy requirement. Then, given that the modified
series is convergent, we could calculate enough coefficients to get an answer 
with the required accuracy. Unfortunately for any other problem than 
the integral, it is difficult to 
calculate the coefficients. A more realistic approach is to 
assume that we can only reach a fixed order and pick the field cutoff in 
such a way that at this order, we reach an optimal accuracy.
Before doing this with a different procedure for even and odd orders as
explained in the introduction, we will first discuss the strong
coupling expansion of Eq. (\ref{eq:int}).

\section{Strong coupling expansion}
\label{sec:sc}

\begin{figure}
\centerline{\psfig{figure=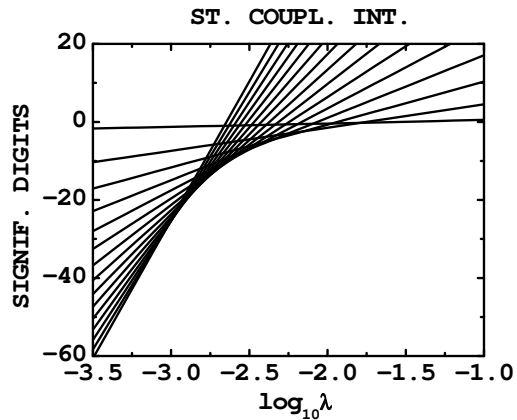,width=3.in}}
\caption{Number of significant digits versus $\lambda$ at order 1, 11,
  21,  ... 141 in the strong coupling expansion. 
As the order increases, the curve rotates counterclockwise and moves
left. }
\label{fig:scenv}
\end{figure}

In the following, we will often use the strong coupling expansion of
the integral (\ref{eq:int}) and a few points should be clarified.
Our original integral $Z(\lambda)$ vanishes in the limit where
$\lambda \rightarrow \infty $. However $\lambda^{1\over 4}Z(\lambda)$
has a finite non-zero limit and can be expanded in powers
$m^2/\lambda^{1/2}$:
\begin{equation}
\lambda^{1\over 4}Z(\lambda)=\sum_{l=0}^{\infty}
(m^2/\lambda^{1/2}))^l b_l\ ,
\label{eq:scexp}
\end{equation}
with 
\begin{equation}
b_l=(-1)^l
(1/2) ^{l+1} (1/l!)
\Gamma (l/2+1/4)
\end{equation}
This expansion is converging over the entire complex plane. However,
if we look at the first few orders displayed later in
Fig. \ref{fig:evenaccs3}, one might be tempted to conclude that the series
has a finite radius of convergence because the curves representing the
significant digits seem to have a ``focus" near $\lambda\ =\
10^{-1.5}$. 
To be completely specific, we mean that in Fig. \ref{fig:evenaccs3},
the four curves labeled S0 to S4 seem to intersect at a given point.
However, as many more orders are displayed, the apparent
focus moves left and a ``caustic'' (envelope) appears. 
This is shown in Fig. \ref{fig:scenv}. 
The only
difference with Fig. \ref{fig:ptenv} is that the region which is
inaccessible is now below. In other words, it is impossible to reach 
arbitrarily low accuracy using the strong coupling expansion!

In lattice field theory, the strong coupling expansion is similar to
the high temperature expansion and for $D>2$ we expect that the
expansion has a finite radius of convergence due to the existence of a
low temperature phase. This difference is not fundamental for the
discussion
which follows because we never use the large order
contributions of the strong coupling expansion. Consequently, 
for any practical purpose, the situation will be similar to the case
where we have a finite radius of convergence.

\section{Even orders}
\label{sec:even}

If $\phi_{max}$ is the only adjustable parameter, the perfect choice
is a solution of:
\begin{equation}
\int_{-\infty}^{+\infty}d\phi \  {\rm e}^{-(m^2/2)\phi^2-\lambda 
\phi^4} =\sum_{k=0}^Ka_k(\phi_{max})\lambda^k\ ,
\label{eq:opteven}
\end{equation}
with 
\begin{equation}
a_k(\phi_{max})={(-1)^k\over{k!}}\int_{-\phi_{max}}^{+\phi_{max}}d\phi \  
{\rm e}^{-(m^2/2)\phi^2}(\phi^4)^k
\end{equation}
Below, we prove that this equation has no solution when $K$ is
odd, and $K$ is assumed to be even in this section.
This equation can be solved numerically
with good accuracy using Newton's  method or a binary search. 
Our goal is to find
approximate methods (which can be used in more complicated situations)
to solve this equation  and compare them with the accurate 
numerical solutions.
In the rest of this section, we consider the cases of 
strong and weak coupling estimates of the optimal 
value of $\phi_{max}(\lambda)$.
\subsection{Strong Coupling Estimates}
\label{subsec:strong}
\begin{figure}
\centerline{\psfig{figure=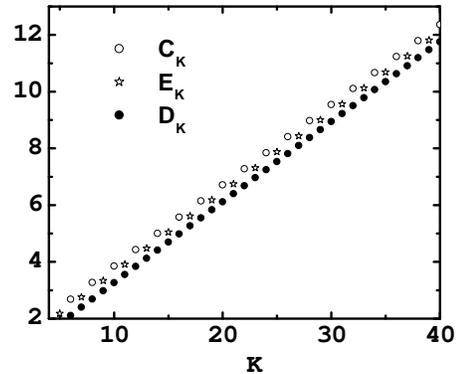,width=3in}}
\caption{Solutions $C^{(0)}_K, D_K$ and $E_K$ defined by
Eqs. (\ref{eq:ck}), 
(\ref{eq:dk}) and, later in the text, by Eq. (\ref{eq:pms}). }
\label{fig:ck}
\end{figure}

\begin{figure}
\centerline{\psfig{figure=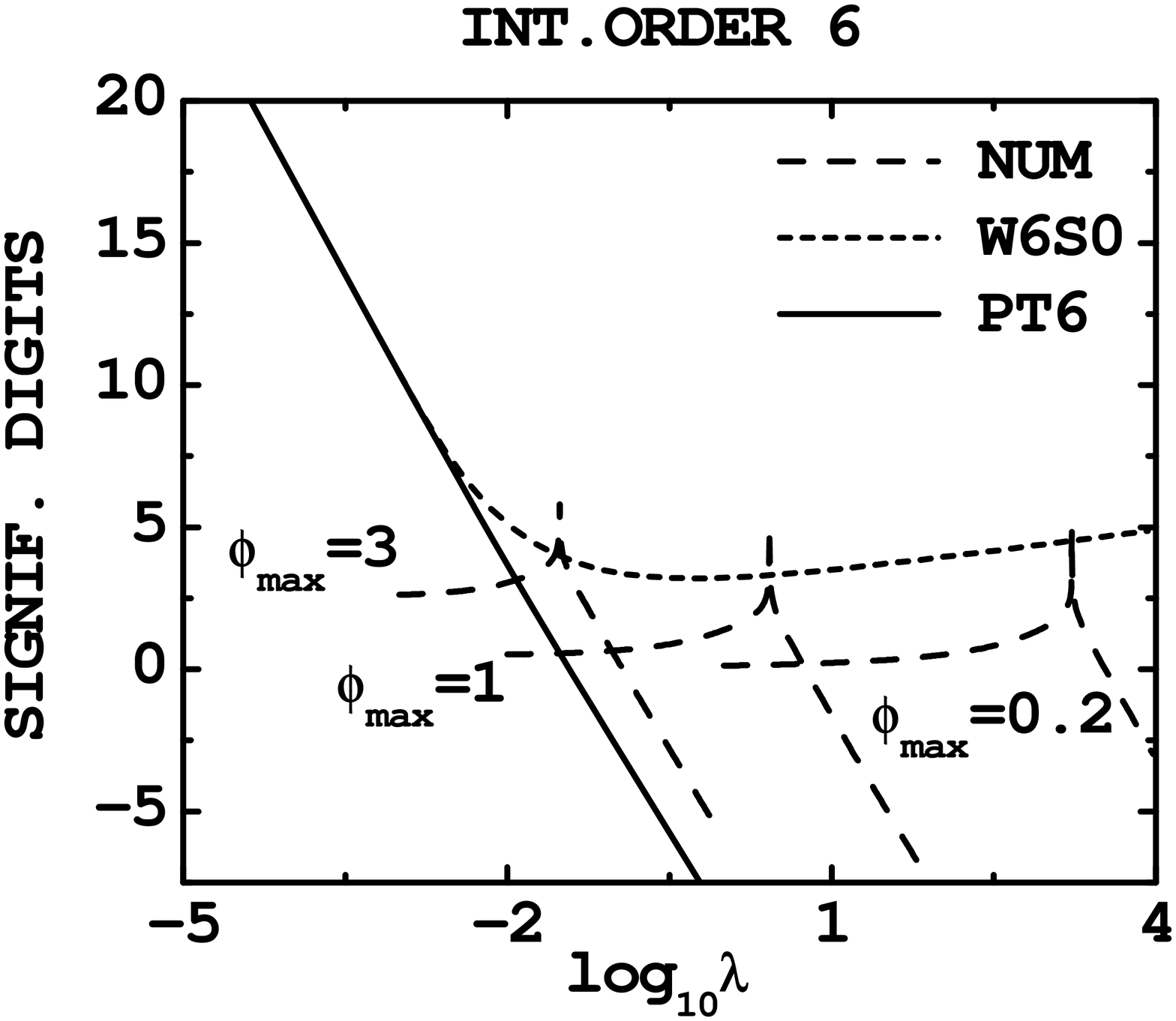,width=3in}}
\caption{Significant digits obtained with the optimal cut 
$\phi_{max}(\lambda)$
estimated using a strong coupling expansion at lowest order (W6S0), 
compared to results at three 
fixed cuts and regular perturbation theory (PT6) at order 6.}
\label{fig:evenaccs0}
\end{figure}

Multiplying both sides of Eq. (\ref{eq:opteven}) by $\lambda^{1/4}$
and expanding in powers of $m^2/\lambda^{1/2}$ we obtain at
zeroth order that $\lambda \phi_{max}^4\simeq C_K^{(0)}$, with
$C_K^{(0)}$ a solution of
\begin{equation}
4\sum_{k=0}^K (-1)^k {(C_K^{(0)})^{k+{1/4}}\over{k!(4k+1)}}=
\Gamma({1/4})\ .
\label{eq:ck}
\end{equation}
The solutions of this transcendental equation are displayed in
Fig. \ref{fig:ck}
for various orders $K$. 
Asymptotically, $C_K^{(0)}\simeq 0.75 + 0.28 K$.
These solutions can be compared with the solutions $D_K$ of the equation
\begin{equation}
{\rm e}^{-D_K}=(D_K)^K/K! \ ,
\label{eq:dk}
\end{equation}
which can be used as a rough estimate of $\lambda
\phi_{max}^4$. Asymptotically,
$D_K\simeq AK+\dots$, where $A=0.278465\dots$ is a solution the
transcendental
equation ${\em e}^{-A-1}=A$.
This lowest order (in the strong coupling) estimate of 
the optimal value of $\phi_{max}$ is quite good.
In Fig. \ref{fig:evenaccs0}, we see that for $K$ = 6 it provides a
significant improvement compared to the regular perturbative series at
order 6 for $\lambda > 10^{-2}$. 
In Fig. \ref{fig:evenaccs0}, we also compare with the accuracy at fixed cuts. 
For a fixed value of $\phi_{max}$, Eq. (\ref{eq:opteven}) has one
solution for a given $\lambda$ and the accuracy becomes infinite at
this value. In the Fig. we see only peaks of finite height, but we see
that the approximation goes quite high in the peak, in other words, we
localize the optimal value quite well. 

\begin{figure}
\centerline{\psfig{figure=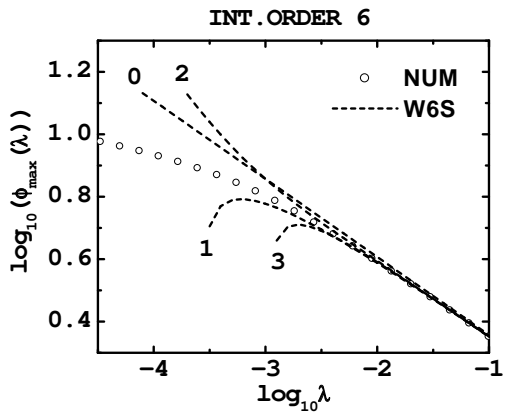,width=3in}}
\caption{Estimates of the optimal $\phi_{max}(\lambda)$ at the first
four orders in the strong coupling expansion }
\label{fig:phiopts}
\end{figure}

\begin{figure}
\centerline{\psfig{figure=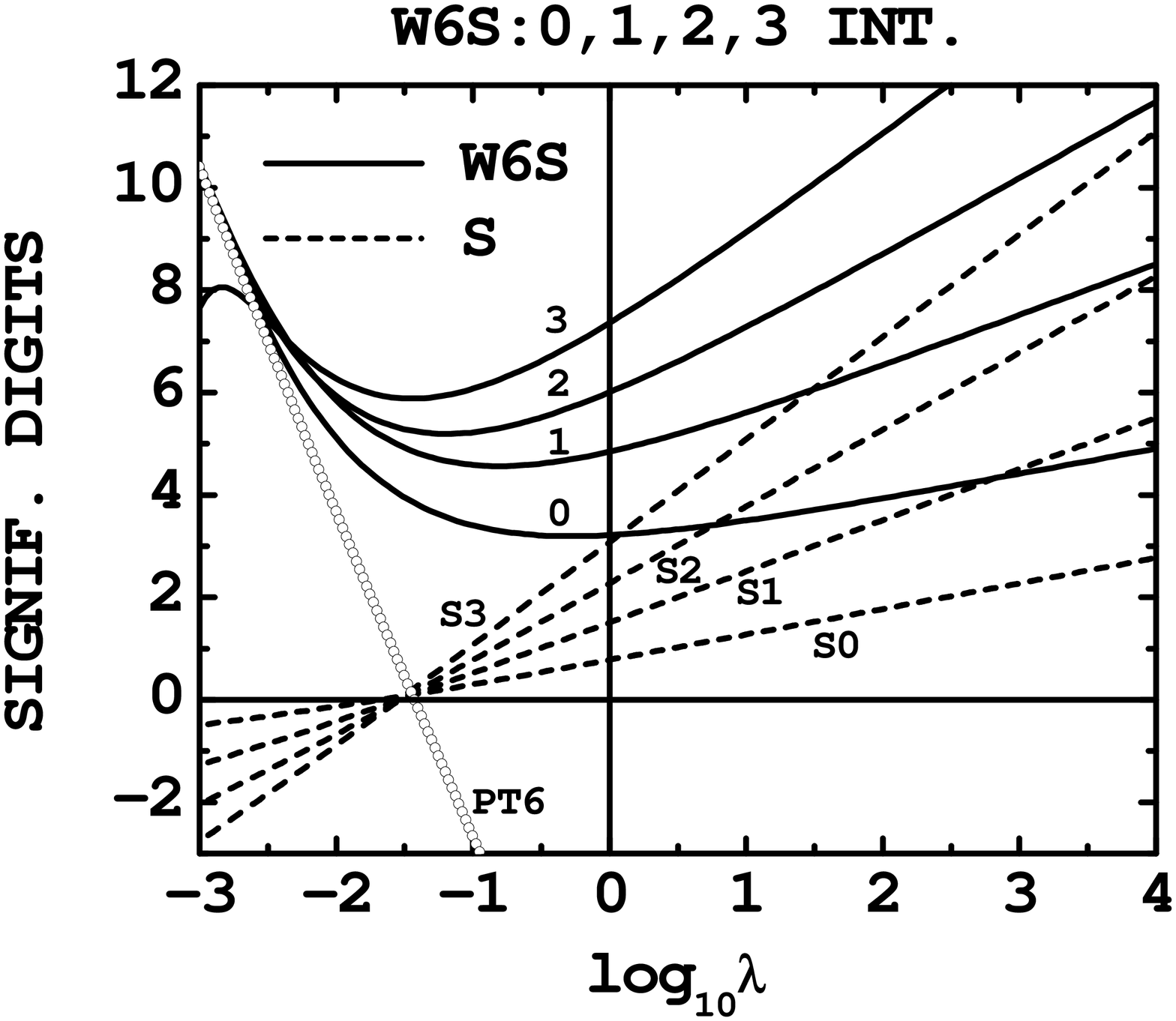,width=3in}}
\caption{Significant digits obtained with the optimal cut 
$\phi_{max}(\lambda)$ (corresponding to a truncated expansion at order
6 in the weak coupling) 
estimated using the strong coupling expansion at orders 0, 1, 2 and 3
(solid lines), 
compared to significant digits using only the strong coupling expansion of
the integral at the same orders in the strong coupling (dashed lines)
and regular perturbation theory at order 6 (PT6).}
\label{fig:evenaccs3}
\end{figure}

We can now proceed to higher orders in  $m^2/\lambda^{1/2}$
using the expansion
\begin{equation}
\lambda \phi_{max}^4 = \sum_{l=0} C_K^{(l)}(m^2/\lambda^{1/2})^l \ ,
\end{equation}
and plug it in the expansion in the same parameter of 
Eq. (\ref{eq:opteven}).
The new coefficients obey linear equations which can be solved order 
by order. The  
optimal $\phi_{max}(\lambda)$ calculated at the four lowest orders in 
 $m^2/\lambda^{1/2}$ are
shown in Fig. \ref{fig:phiopts}. As explained in section \ref{sec:sc},
below a certain value of $\lambda$ a few orders in the strong coupling
expansion won't help and one
needs much higher order to improve the estimate in this region. 
After a short reflection, one can conclude that the ``focus'' observed
in Fig. \ref{fig:evenaccs3} is compatible with Fig.  \ref{fig:phiopts}.

A few words should be said about the notations we use for the curves
in the figures. When we write W6S1,
this means that we use the weak coupling expansion up to order $K$=6
in Eq. (\ref{eq:opteven})(this
is the W6 part) and a strong coupling expansion at order 1 in 
$m^2/\lambda^{1/2}$ (this is the S1 part) in the calculation of the 
optimal $\phi_{max}$. In addition, PT8 means the 8th order in regular 
perturbation theory. In some figures, some of the indexes appear
directly near the corresponding curve.

The accuracy of the truncated series
at $\phi_{max}$ calculated
with the higher order corrections in $m^2/\lambda^{1/2}$ 
in Fig \ref{fig:evenaccs3}. 
For comparison, the accuracy obtained by using only the
strong coupling expansion Eq. (\ref{eq:scexp}).
is also shown. The figure makes clear that the method
proposed here represents a significant improvement compared to the
separate use  of the conventional weak and strong coupling expansions.

\subsection{Weak coupling}
\label{subsec:weak}

As we learned in the previous subsection, as $\lambda$ decreases, the
optimal value of $\phi_{max}$ increases. In the limit of a weak
coupling, the ``tails'' of the integral that we removed become 
a small quantity. It is thus advantageous to split the l.h.s. of Eq. 
(\ref{eq:opteven}) into its bulk and tails and expand ${\rm e}^{-\lambda 
\phi^4}$ in the bulk where it is justified. The resulting (exact) equation
for $\phi_{max}$ is then

\begin{equation}
2 \int_{\phi_{max}}^{+\infty}d\phi \  {\rm e}^{-(m^2/2)\phi^2-\lambda 
\phi^4} =-\sum_{k=K+1}^{\infty}a_k(\phi_{max})\lambda ^k \ ,
\label{eq:wopteven}
\end{equation}
In the limit of very small $\lambda$, Eq. (\ref{eq:wopteven}) becomes
\begin{equation}
(2/(m^2\phi_{max})){\rm e}^{-(m^2/2)\phi_{max}^2}\simeq
-a_{K+1}\lambda^{K+1}\ .
\label{eq:wapp1}
\end{equation}
The l.h.s. has a functional form similar to semi-classical estimates
of the energy shifts in quantum mechanics \cite{ym02,latt02}.
A more refined version of this equation is 
\begin{equation}
2 \int_{\phi_{max}}^{+\infty}d\phi \  {\rm e}^{-(m^2/2)\phi^2-\lambda 
\phi^4} \simeq-a_{K+1}(\phi_{max})\lambda ^{K+1} \ ,
\label{eq:wapp2}
\end{equation}
It is clear that the two above equations have solutions only when $K$
is even because in this case $a_{K+1}<0$. In Appendix A, it is shown that
this property extends to the exact Eq. (\ref{eq:opteven}).
More precisely the r.h.s. of Eq. (\ref{eq:opteven}) is positive
for $K$ even and negative for $K$ odd. 

\begin{figure}
\centerline{\psfig{figure=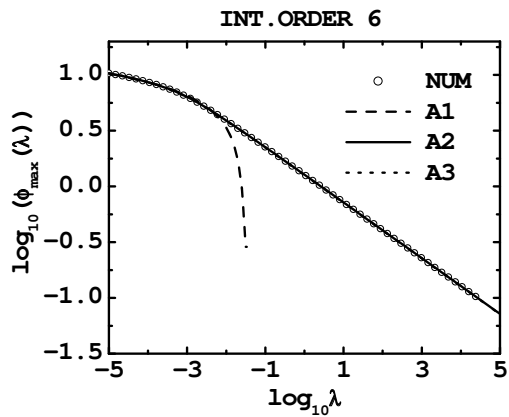,width=3in}}
\caption{Estimates of the optimal $\phi_{max}(\lambda)$ obtained with 
the approximations $A1$, $A2$ and $A3$ defined in the text for $K$ = 6, 
compared to numerical values (empty circles).}
\label{fig:phimaxw}
\end{figure}

\begin{figure}
\centerline{\psfig{figure=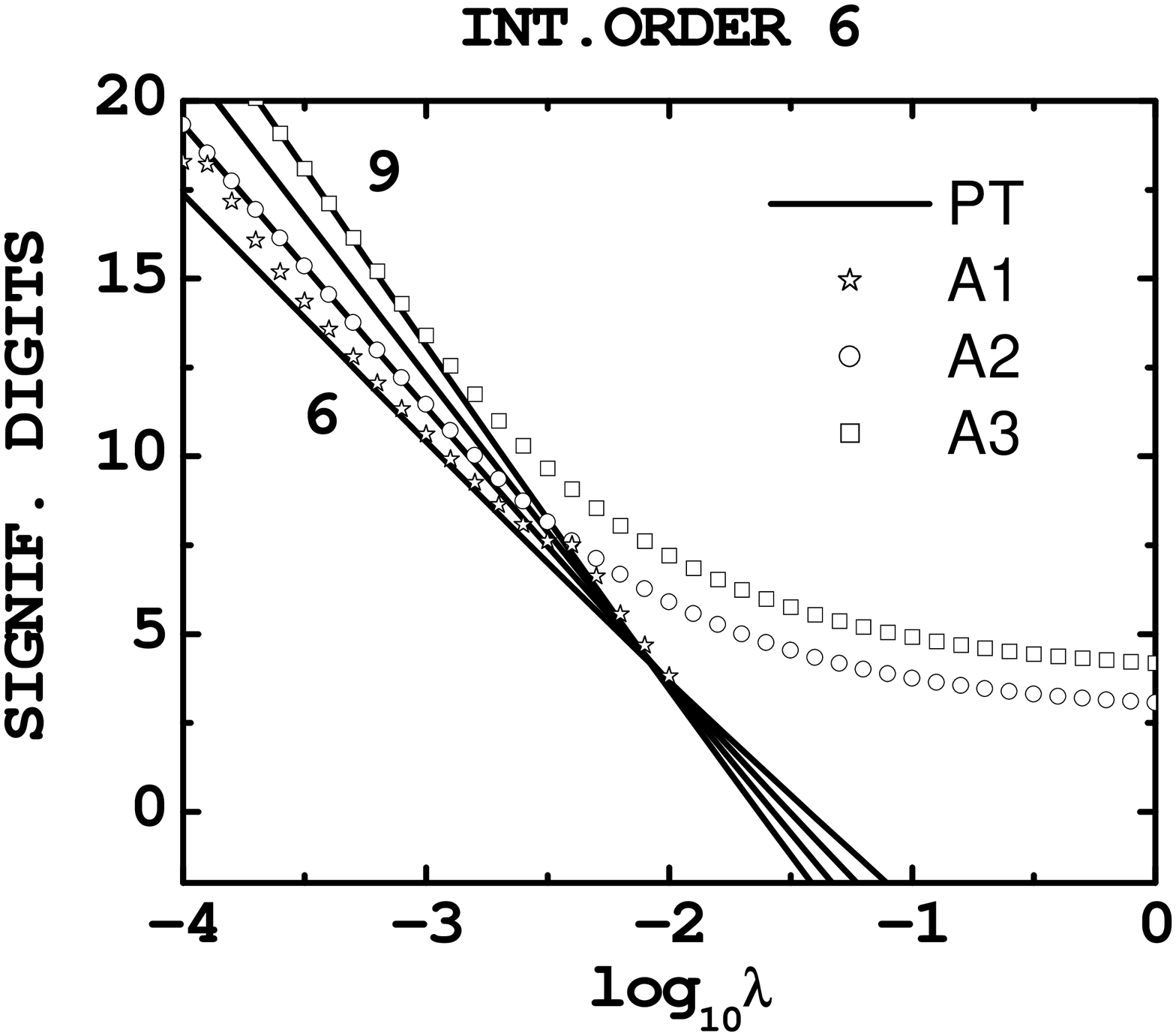,width=3in}}
\caption{Significant digits obtained with the 
approximations $A1$, $A2$ and $A3$ for $K$ = 6,  compared to order 6
to 9 in regular perturbation theory (solid lines rotating clockwise as
the order increases).}
\label{fig:evenwacc}
\end{figure}

Eq. (\ref{eq:wapp2}) can be further improved by including higher order 
truncations at odd orders:
\begin{equation}
2 \int_{\phi_{max}}^{+\infty}d\phi \  {\rm e}^{-(m^2/2)\phi^2-\lambda 
\phi^4} =-\sum_{k=K+1}^{K+3}a_k(\phi_{max})\lambda ^k \ ,
\label{eq:wapp3}
\end{equation}
and so on. In the  following, we refer to the successive
approximations defined by Eqs. (\ref{eq:wapp1}), (\ref{eq:wapp2}) and 
(\ref{eq:wapp3}) as approximations $A1$, $A2$ and $A3$ respectively.
The estimates of the optimal $\phi_{max}$ obtained with these
approximations and the corresponding accuracies as a function of the 
coupling are shown in Figs. \ref{fig:phimaxw} and \ref{fig:evenwacc}.
One can see that A1 provides good estimates of $\phi_{max}$ optimal
only at very small $\lambda$. On the other hand, A2 and A3 both provide good
estimates 
even at large coupling. Not surprisingly, the accuracy of A2 (A3)
merges with order 7 (9) in regular perturbation theory.

\section{Odd orders}
\label{sec:odd}

\begin{figure}
\centerline{\psfig{figure=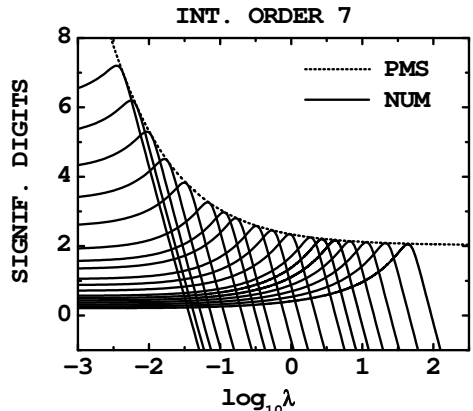,width=3in}}
\caption{Significant digits obtained with the PMS condition 
compared to the same quantity at
fixed cuts.}
\label{fig:pms}
\end{figure}

As explained in section \ref{sec:even}, 
for series truncated at odd $K$, the r.h.s. of
Eq. (\ref{eq:opteven}) is negative. The best that we can do is to
minimize the error (i. e., the difference between the r.h.s. and the
l.h.s. . The minimization condition implies that  $\lambda
\phi_{max}^4 = E_K$, the unique (see Appendix \ref{app:evenodd},  
where $f_K$ is defined) 
solution of 
\begin{equation}
\sum_{k=0}^K{(-E_K)^k \over{k!}}=f_K(E_k)=0 \ .
\label{eq:pms}
\end{equation}
In the following, we refer to this condition as the Principle of
Minimal Sensitivity (PMS) condition. 
This terminology has been used \cite{buckley93} 
in the LDE where the variational parameter is fixed 
by requiring that the final estimate depends as least as possible 
on this parameter.
The solutions $E_K$ are displayed
in Fig. \ref{fig:ck} which shows that they are asymptotically close to
the solutions $C_{K-1}^{(0)}$ obtained at the lowest order in the strong 
coupling expansion. The accuracy obtained using the PMS condition is 
by construction the envelope of the accuracy obtained for all possible
$\phi_{max}$. This is illustrated in Fig. \ref{fig:pms}.

\begin{figure}
\centerline{\psfig{figure=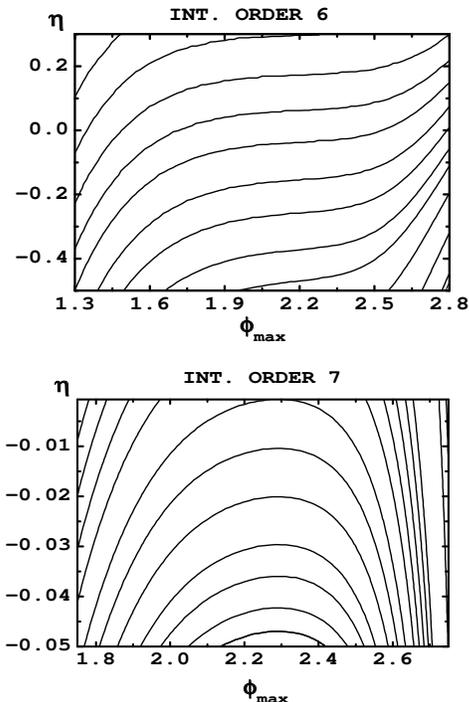,width=2.9in}}
\caption{Level curves of the perturbative expansion of 
$Z(\lambda ,\phi_{max},\eta)$ 
at order 6 and 7 in $\lambda$ evaluated at $\lambda=0.1$.}
\label{fig:level}
\end{figure}

Nevertheless, an exact match between the original integral and the
truncated perturbative expansion with a field cut  
can be obtained by using a mass
shift $m^2\rightarrow m^2(1+\eta)$. Using obvious notations, we denote
the cut integral defined in Eq. (\ref{eq:cutint}) with this mass shift
$Z(\lambda ,\phi_{max},\eta)$. The level curves of the perturbative 
expansion of $Z(\lambda ,\phi_{max},\eta)$ at fixed $\lambda$
follow different patterns at odd and even orders as
illustrated in Fig. \ref{fig:level}. 

\begin{figure}
\centerline{\psfig{figure=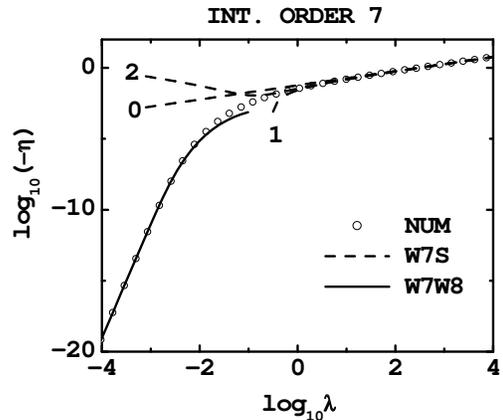,width=3in}}
\caption{Optimal $\eta$ as a function of $\lambda$. Numerical values 
(empty circles) compared to approximate values obtained with 
the weak coupling expansion expansion (W7W8) corresponding to
Eq. (\ref{eq:w8}) with $K=7$ and the orders 0, 1 and 2 in
the strong coupling expansion (W7S) from Eq. (\ref{eq:etasc}).}
\label{fig:etacomp}
\end{figure}

\begin{figure}
\centerline{\psfig{figure=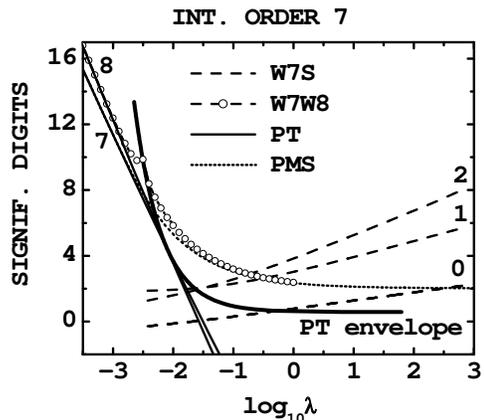,width=3in}}
\caption{Significant digits for the same approximations as in 
Fig.\ref{fig:etacomp} compared to the regular perturbative series at 
orders 7 and 8 (thin solid line), the PMS result with $\eta$=0 at order 7 
(small dots)
and the envelope of
regular perturbation theory (thick solid line). }
\label{fig:sdeta}
\end{figure}

\begin{figure}
\centerline{\psfig{figure=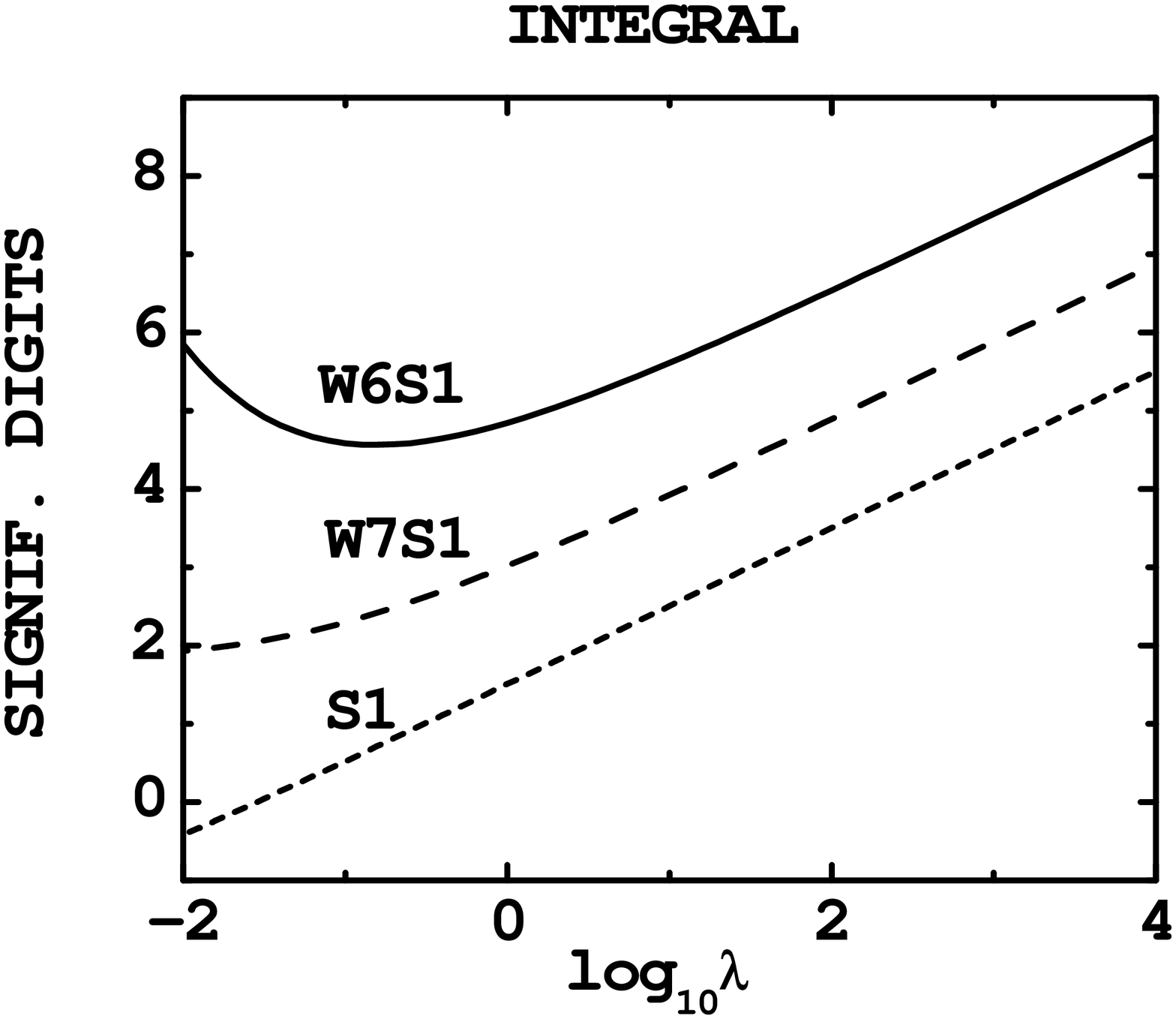,width=3in}}
\caption{Significant digits obtained with the strong coupling
expansion at order one in  $m^2/\lambda^{1/2}$ and the approximations
W6S1 and W7S1 discussed previously. }
\label{fig:scompa}
\end{figure}

At even order, all the level curves cross the $\eta =0$ line and there
is no need for a mass shift. This case was discussed in section
\ref{sec:even}.
At odd order, the level curve
corresponding to the exact value $Z(\lambda)$ defines 
a curve  $\eta(\phi_{max})$. In appendix \ref{app:eta}, we show that
this curve stays in the half-plane $\eta<0$.  
We are now free to pick an arbitrary  value of $\phi_{max}$ and adjust
$\eta=\eta(\phi_{max})$. In the following we will pick $\phi_{max}$ in
such a way that $\eta$ is as small as possible. This can be
accomplished by solving the equation
\begin{equation}
\partial
\eta/\partial \phi_{max} =0\ ,
\end{equation}
for $\phi_{max}$.
In appendix \ref{app:eta}, we show that this condition is indeed
equivalent to the PMS condition (\ref{eq:pms}) and consequently we have
simply $\phi_{max}=(E_K/\lambda)^{1/4}$. With this choice, the
introduction of $\eta$ is a natural continuation of the optimization
at $\eta$ =0.
Estimations of $\eta$  for this choice of $\phi_{max}$ can be obtained 
approximately at strong and weak coupling.
In the limit of arbitrarily 
small coupling, we can treat $\eta$ as a quantity of order
$\lambda^{K+1}$ and use it to make up for the ``missing'' even
contribution that would allow a solution of Eq. (\ref{eq:opteven}).
This reasoning implies the weak coupling estimate
\begin{equation}
\eta\simeq-\lambda^{K+1}a_{K+1}(\phi_{max})m(2/\pi)^{1/2}
\label{eq:w8}
\end{equation}
On the other hand, 
at strong coupling, $\eta$ grows like $\lambda^{1/2}$ and we need to
expand
\begin{equation}
m^2\eta\lambda^{-1/2}=\sum_{l=0}^{\infty}B_l(m^2/\lambda^{1/2})^l \ .
\label{eq:etasc}
\end{equation}
The two approximation work well in their respective range of validity 
as shown in Fig. \ref{fig:etacomp}. 
The significant digits obtained with the various procedures are
displayed in Fig. \ref{fig:sdeta}. One sees that 
the mass shift provides a significant improvement compared to the
PMS condition at $\eta$=0. 
If we compare the two methods in their respective region of validity,
the improvement provided by the strong 
coupling method is more substantial. Not surprisingly, W7W8 merges
with PT8 at weak coupling.
We can now compare the accuracy of various estimates based on strong 
coupling expansions at the
same order in $m^2/\lambda^{1/2}$. Examples are shown in 
Fig. \ref{fig:scompa} where the accuracy obtained with three methods 
relying on estimates at order one in $m^2/\lambda^{1/2}$ are displayed.
One can see that as the coupling becomes large, the
accuracy increases at the same rate in the three cases. As we already
know in the even case, our method significantly improves the basic
strong coupling expansion from Eq. (\ref{eq:scexp}). However, 
the improvement based on even order $K=2q$ in $\lambda$ 
performs significantly better than the improvement based on the odd
order $K=2q+1$. Consequently, when in the next section we compare with
other existing methods, we will restrict ourselves to the even case.
\section{comparison with other methods}
\label{sec:compa}
\begin{figure}
\centerline{\psfig{figure=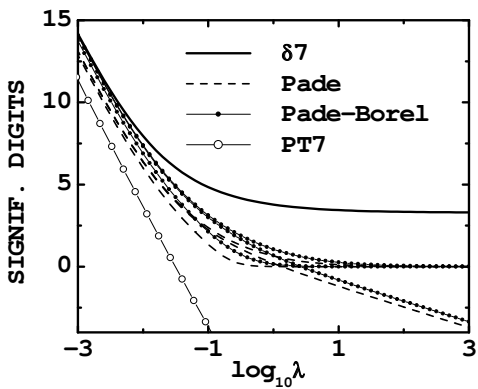,width=3in}}
\caption{Comparison of the delta expansion at order 7 ($\delta 7$) with Pad\'e
approximants and the Pad\'e-Borel method. In both cases the
approximants [4/3], [3,4] and [2,5] have been used.}
\label{fig:padecompa}
\end{figure}

\begin{figure}
\centerline{\psfig{figure=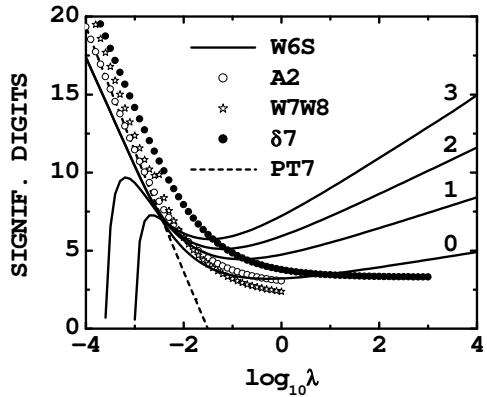,width=3in}}
\caption{Comparison of the delta expansion at order 7 ($\delta 7$) 
with various
methods discussed previously and regular perturbation at order 7 (PT7).}
\label{fig:summary}
\end{figure}

There exists several methods to improve the accuracy of asymptotic
series. These include Pad\'e's approximants \cite{baker96} applied to 
the series itself or its Borel-transform and the LDE \cite{buckley93}. 
These methods are compared among
themselves in Fig. \ref{fig:padecompa}.
One can see that at weak coupling, the LDE provides an
upper envelope for the accuracy while at strong coupling it prevails
more significantly. Consequently, we only need to compare our results 
to the LDE. This is done in
Fig. \ref{fig:summary} 
where we see that at strong and moderate coupling, our methods provide
a significant improvement compared to the LDE.
On the other hand, at weak coupling, the improvements that we
proposed, do not perform as well as the LDE. 

The results of Fig. \ref{fig:summary} have been obtained by
making the replacements \cite{buckley93} 
$m^2\rightarrow\Omega^2+\delta(m^2-\Omega^2)$
and $\lambda\rightarrow \delta \lambda$. We then expanded the 
perturbative series at order $K$ in $\lambda$ to order $K$ in $\delta$.
The arbitrary
parameter $\Omega^2$ was determined by
requiring that the derivative of the estimate with respect to $\Omega^2$ 
vanishes. This was called the PMS condition in
Ref. \cite{buckley93} and it has a solution at odd orders only.

\begin{figure}
\centerline{\psfig{figure=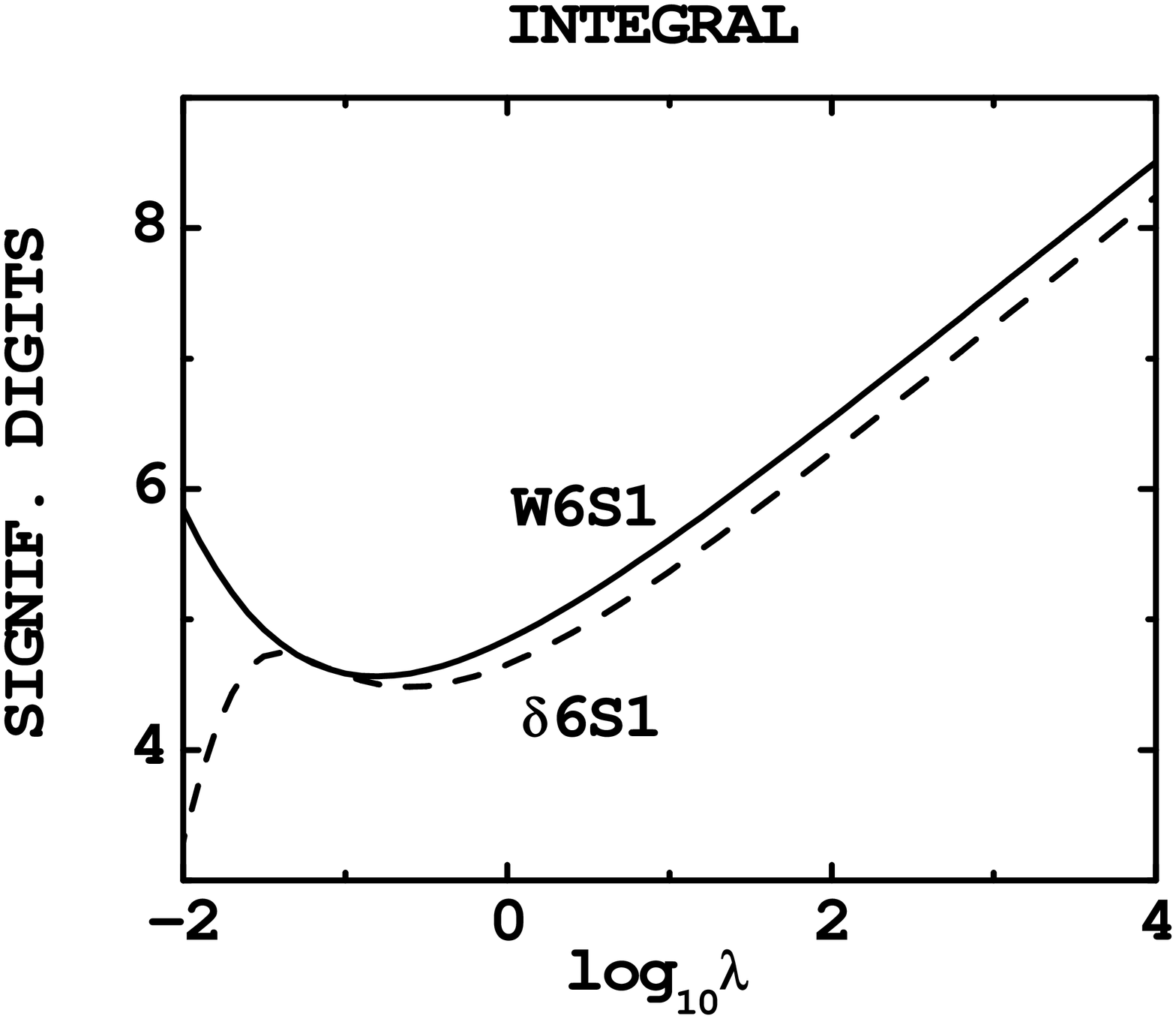,width=3in}}
\caption{Significant digits obtained using the strong coupling
expansion at order one in  $m^2/\lambda^{1/2}$ to determine $\Omega$
in the $\delta$ expansion at order 6 compared 
the approximation
W6S1 discussed previously. }
\label{fig:dels}
\end{figure}

At even orders, it is however possible to proceed in a way similar to
what we have done in subsection \ref{subsec:strong}, namely matching
the strong coupling expansion of the estimate with the usual strong
coupling of the integral in order to determine the arbitrary
parameter. At order zero in $m^2/\lambda^{1/2}$, this results into a
transcendental equation which has a solution at even orders only. 
Higher orders corrections to $\lambda/\Omega^4$ can then be calculated
by solving linear equations just as in subsection \ref{subsec:strong}.
The numerical results for $K=6$ are displayed in Fig. \ref{fig:dels}.
One can see that this method and the method presented in subsection 
\ref{subsec:strong} have very similar accuracy at 
moderate and strong coupling. 

The procedure we have used above is closely related to variational
methods \cite{var} where weak and strong coupling expansions 
were combined for various purposes. The only difference is that here 
we simply imposed the matching with the strong coupling expansion
rather than resorting to extremization procedures or large order
scaling arguments.

\section{Conclusions}

In conclusion, for even series in the weak coupling, the 
method which consists in determining the optimal value of $\phi_{max}$ 
using the strong coupling expansion provides excellent results 
at moderate at strong coupling. There is room for improvement at weak
coupling. In particular, progress could be made by finding accurate
approximations to calculate a large number of terms in 
the r.h.s. of Eq. (\ref{eq:wopteven}).

The methods used here can be extended to quantum mechanics and in
particular for the anharmonic oscillator where similar calculations 
have been partially performed\cite{progress}.
We are planning to apply the methods developed here for higher
dimensional SFT where the LDE seems to converge very slowly
\cite{braaten} (see also 
\cite{kneur,kleinert} for methods 
to improve the situation). One difficulty is to calculate the
perturbative coefficients with a field cutoff. Monte Carlo methods 
have been recently developed for this purpose \cite{tsukuba}.

This research was supported in part by the Department of Energy
under Contract No. FG02-91ER40664. 
Y. M. was at the Aspen Center for
Physics while this work was in progress. 
Y. M. was partly supported by a 
Faculty Scholar Award at The University of Iowa and a residential
appointment at the Obermann Center for Advanced Studies at the
University of Iowa, while the manuscript was written.

\appendix

\section{Non-existence of solutions in the odd case}
\label{app:evenodd}
In this appendix, we show that Eq. (\ref{eq:opteven}) has no solution
when $K$ is odd. For this purpose we introduce the truncated
exponential series:
\begin{equation}
f_K(x)=\sum_{k=0}^K (-1)^k x^k/k! \ ,
\end{equation}
and their complement
\begin{equation}
g_K(x)=\sum_{k=K}^{\infty}(-1)^k x^k/k! \ .
\end{equation}
Using the fact that $f'_{K}=-f_{K-1}$ and a similar relation for the
$g$, one can show by induction that for $K$ even, $f_K$ is strictly
positive and $g_K$ is positive with its only zero at zero. 
For $K$ odd and $x>0$, $g_K$ is negative and decreases. 
Given that the r.h.s. of Eq. (\ref{eq:wopteven}) is the integral
with a positive measure of $-g_{K+1}$ over positive argument, we see
that the r.h.s. is positive when $K$ is even and negative when $K$
is odd. Since the l.h.s. is always positive, they are no solutions
for $K$ odd.

\section{Special Features of $\eta(\phi_{max})$}
\label{app:eta}

The function $\eta(\phi_{max})$ is the solution of the equation
\begin{equation}
Z(\lambda)=\int_{-\phi_{max}}^{\phi_{max}}{\rm e}^
{-(m^2/2)(1+\eta(\phi_{max}))\phi^2}f_K(\lambda \phi^4)\ .
\label{eq:etaphi}
\end{equation}
In this appendix, $K$ is assumed to be odd.
For $x>0$, we have $f_K(x)< {\rm e}^{-x}$ because
$g_{K+1}(x)>0$ for $K+1$ even and $x>0$ (see appendix
\ref{app:evenodd}).
We can compensate this underestimation by making the integration
measure more positive, in  other words picking the parameter $\eta<0$.
The l.h.s. of Eq. (\ref{eq:etaphi}) is independent of $\phi_{max}$. 
Taking the derivative of Eq. (\ref{eq:etaphi}) with respect to
$\phi_{max}$ 
and imposing that  $\phi_{max}$ is a solution of $\partial
\eta/\partial \phi_{max} =0$ ,
we obtain that for this special value of $\phi_{max}$, we have 
\begin{equation}
f_K(\lambda \phi_{max}^4){\rm
  e}^{-(m^2/2)(1+\eta(\phi_{max}))\phi_{max}^2}=0\ .
\end{equation}
which implies the PMS condition (\ref{eq:pms}).

\end{multicols}
\end{document}